\documentclass[a4paper,fleqn]{cas-dc}

\usepackage[numbers]{natbib}

\providecommand{\botrule}{\bottomrule}

\usepackage{float}
\restylefloat{figure}
\usepackage[section]{placeins}

\begin{document}
\let\WriteBookmarks\relax
\def\floatpagepagefraction{1}
\def\textpagefraction{.001}

\shorttitle{VPhC Biosensor for RI Discrimination of Cancer Cell Lines}

\shortauthors{J. Hossain et~al.}

\title[mode=title]{Harnessing Topological Valley-Hall States in Photonic
Crystals for Label-Free Refractive-Index Discrimination of Cancer Cell Lines}

\author[1]{Junayet Hossain}
\ead{junayethossain260@gmail.com}

\author[1]{Sajid Muhaimin Choudhury}

\author[1]{Mohammed Imamul Hassan Bhuiyan}
\cormark[1]
\ead{imamul@eee.buet.ac.bd}

\affiliation[1]{organization={Department of Electrical and Electronic
                Engineering, Bangladesh University of Engineering and
                Technology},
                city={Dhaka},
                postcode={1205},
                country={Bangladesh}}

\cortext[cor1]{Corresponding author}

\begin{abstract}
Topological photonic crystals (TPhCs) are the photonic analogs of topological
insulators, inspired by the quantum Hall effect and its variants, including the
integer, spin, and valley Hall effects. Among these, valley-Hall photonic
crystals (VPhCs) achieve topological protection by breaking inversion symmetry,
making them closely related to valleytronics. Like other TPhCs, VPhCs enable
robust, backscattering-free light propagation, even in the presence of
structural imperfections, sharp bends, or defects. In this study, we introduce a
silicon VPhC design near 6~THz that demonstrates exceptional resilience in
guiding light through both linear and $\Omega$-shaped waveguides
($-1.12$ and $-1.27$~dB insertion loss), even under significant structural
disorder ($0.47$~dB worst-case degradation). Additionally, efficient light
confinement is demonstrated within a hexagonal resonant cavity, further
underscoring the versatility of the platform. On this basis, a VPhC-based
biosensor is proposed, designed to achieve high sensitivity and a strong quality
factor, thereby enabling carcinoma cell detection ($\Delta n = 0.011$~RIU). The
work demonstrates that the sensor achieves a maximum quality factor of
285{,}338, a maximum sensitivity of 24{,}300~nm/RIU, and a figure of merit of
132{,}813~RIU$^{-1}$. This work holds significant promise for advancing medical
and clinical diagnostics, paving the way for innovative applications in
healthcare.
\end{abstract}



\maketitle

\section{Introduction}\label{sec:intro}

The Quantum Hall Effect bridged mathematical topology and solid-state
physics, unlocking non-trivial topological phases such as
topological insulators \cite{moore2010birth}, Chern insulators
\cite{PhysRevB.74.235111}, and topological superconductors
\cite{sato2017topological}. Topological insulators host conductive edge
states while remaining insulating in the bulk, a property guaranteed by a
finite bandgap
\cite{kane2005z,katmis2016high,kane2005quantum,zhang2009topological,li2016experimental,christensen2022location},
and the same principles extend to photonics and optical systems
\cite{PhysRevA.84.043804,khanikaev2013photonic,hafezi2013imaging,rechtsman2013photonic}.
When materials of distinct topological character are brought into contact, a robust edge state
appears at the interface. The first topological photonic insulator used
gyromagnetic material with time-reversal symmetry broken by a magnetic field
\cite{wang2009observation}; subsequently, breaking spatial symmetry realized
bosonic analogs of the spin-Hall and valley-Hall effects in metallic
\cite{ma2015guiding,chen2014experimental,wu2017direct} and dielectric
materials \cite{wu2015scheme,chen2017valley,dong2017valley}, avoiding the weak
magnetic response of optical materials
\cite{PhysRevLett.100.013904,PhysRevA.78.033834}. The valley-Hall effect, a
valleytronic analog, has likewise been realized \cite{liu2021valley}. Because
valley degrees of freedom are defined in momentum space, edge states require an
interface between regions of distinct topological invariants
\cite{xiao2007valley}. Applications of topologically protected edge
states include information encoding \cite{si2024topological}, sensing
\cite{kumar2022topological}, and photonic logic \cite{he2019topological}.
The terahertz (THz) range is well suited to integrated sensing: it contains
characteristic absorption bands of many materials and molecules---explosives
\cite{chen2007absorption,liu2006detection}, proteins \cite{niessen2019protein},
and DNA \cite{fischer2002far}---and is non-ionizing, so it does not damage
biological tissue \cite{pickwell2006biomedical,macpherson2013guest}. Sensor
efficiency is governed by light--matter interaction, enhanced either by a longer
interaction length or a high-$Q$ resonant cavity. Metamaterials---aperiodic
arrays of subwavelength resonators---are a common strategy
\cite{smith2004metamaterials,smith1999loop,xu2017mechanisms}, valued for their
simple structure and tight field confinement
\cite{chen2012membrane,o2008thin,singh2014ultrasensitive}, but their $Q$ is
limited by metallic loss and they integrate poorly with silicon. Waveguide-based
sensors \cite{zhang2004waveguide,mendis2009terahertz} instead extend the
effective interaction length with the analyte. Recent photonic devices exploiting the energy-valley degree of
freedom, based on valley photonic crystals (VPhCs), have advanced topological
control of light. Topologically robust transport was demonstrated
in a silicon-on-insulator (SOI) VPhC slab along sharp-bend interfaces at
subwavelength scales \cite{he2019silicon}. In 2020, a chiral nanophotonic
resonator exploiting edge states at the interface of distinct VPhCs was proposed
\cite{barik2020chiral}. A chiral quantum-optical interface was later realized
by integrating semiconductor quantum dots into a VPhC waveguide
\cite{jalali2020chiral}. Building on these demonstrations, topological photonic
crystal cavities have been explored for THz sensing. Kumar et al.\
\cite{kumar2022topological} demonstrated an on-chip silicon VPhC sensor
achieving $Q = 1.4\times10^5$ for thin-film detection, while Navaratna et al.\
\cite{navaratna2023chip} reported Q-factors approaching $2.0\times10^5$ in the
0.3--0.6~THz band for solvent and hydration sensing. Both designs rely on
evanescent field interactions at cavity boundaries, limiting the achievable
light--matter coupling strength.
Topological refractive-index sensing has advanced rapidly in the past two
years. Li et al.\ \cite{li2025vpc} reported the first valley-photonic-crystal
(VPhC) Mach--Zehnder RI sensor (intensity sensitivity 1534\%/RIU at
1.53~$\mu$m, CMOS-compatible), and Satyaraj et al.\ \cite{satyaraj2024}
demonstrated a VPhC ring-resonator biosensor for mid-infrared brain-tumor
discrimination ($9021$~nm/RIU, $Q = 2.9\times10^4$). In parallel,
one-dimensional Zak-phase designs have reached high figures of merit:
Dash et al.\ \cite{dash2025} reported a graded-index topological resonator
for cancer cell-line sensing ($1806$~nm/RIU, FoM $= 4030$~RIU$^{-1}$),
Anjineya et al.\ \cite{anjineya2025} a Zak-phase interface sensor
(FoM $\approx 2.9\times10^4$~RIU$^{-1}$), and Barvestani \cite{barvestani2025}
a porous-cap interface-state biosensor for aqueous analytes; the field is
surveyed in \cite{abood2025}. These works operate at telecom-to-mid-infrared
wavelengths and rely on evanescent or interface-mode field overlap; the
application of \emph{valley-Hall} topological cavities to THz cancer
cell-line discrimination with volumetric (rather than evanescent) analyte
loading remains unexplored. In this work, we propose a VPhC-based sensor that addresses this gap through
three specific contributions: (i)~demonstration of
robust topological waveguiding in both linear and $\Omega$-shaped geometries
with worst-case transmission degradation of only 0.47~dB under structural
defects (power above 73\% retained in the $\Omega$-geometry)---deterministic defect tolerance under simultaneous sharp bends and localized
perturbations in a biosensor-relevant geometry that, to our knowledge, has not
been quantified in this form; (ii)~formation of a hexagonal topological cavity with volumetric
analyte loading ($\Gamma = 0.68$ versus $<$0.20 for evanescent
designs \cite{kumar2022topological}) and quality factors up to
285,338, exceeding Navaratna et al.\ \cite{navaratna2023chip} by more than
42\% within the present 2D simulation framework, while targeting cancer
cell-line discrimination rather than solvent
sensing; and (iii)~proof-of-concept refractive-index discrimination of five
cancer-derived cell lines---Jurkat, HeLa, PC-12, MDA-MB-231, and
MCF-7---with a maximum sensitivity of 24,300~nm/RIU and a maximum figure of
merit of 132,813~RIU$^{-1}$.

\section{Design of the VPhC}\label{sec:vphc}

\begin{figure}[H]
    \centering
    \includegraphics[width=\linewidth]{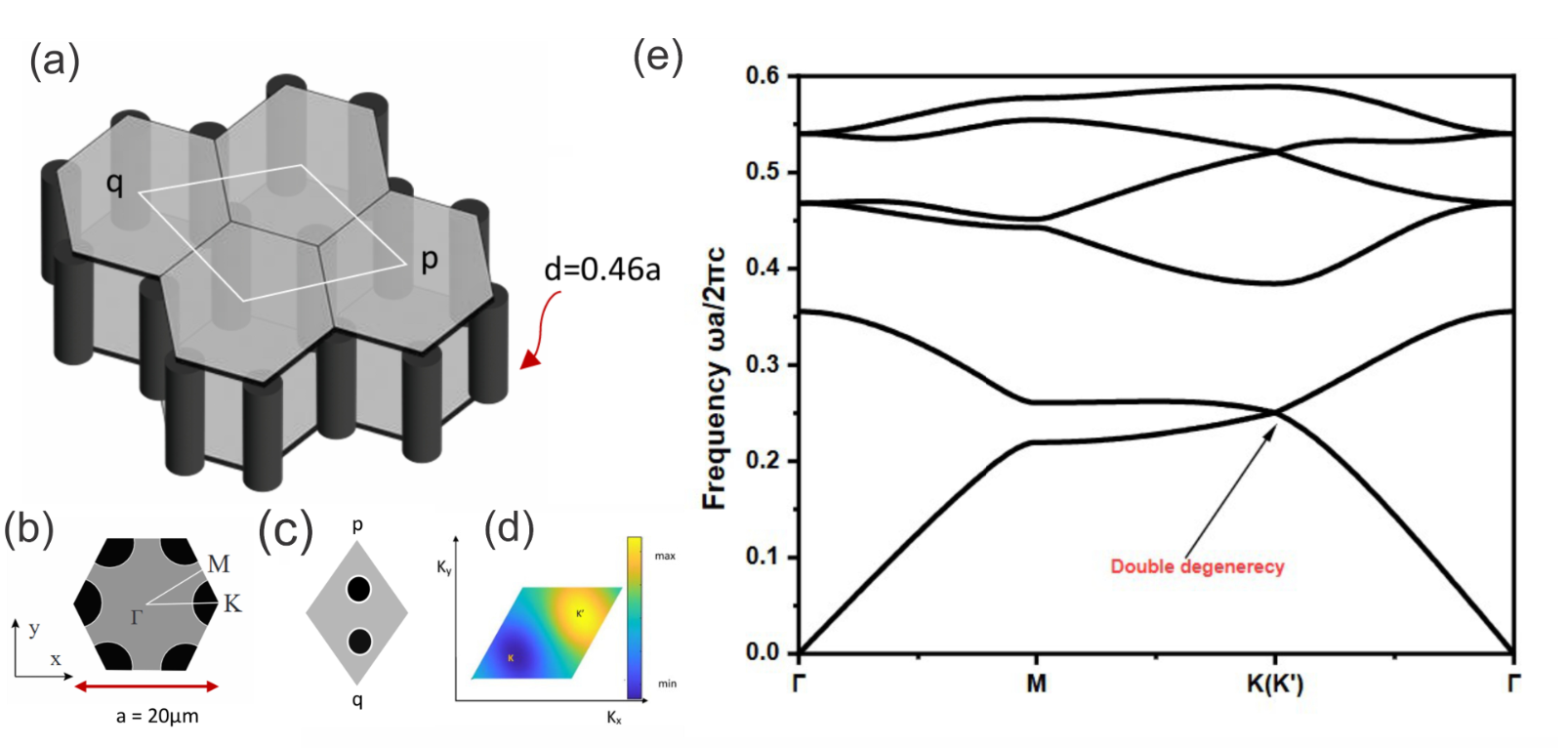}
    \caption{\textbf{Design of the primary VPhC.} (a) Honeycomb lattice with rod
    diameter $d = 0.46a$. (b) Top view of a single unit cell, lattice constant
    $a = 20~\mu$m. (c) Rhombic representation. (d) Berry curvature distribution
    with opposite signs at K and K$'$, compatible with $C_K = \pm 1/2$.
    (e) Band structure illustrating the Dirac point and double degeneracy
    at K/K$'$.}
    \label{fig:1}
\end{figure}

The valley-Hall photonic crystal is constructed from silicon rods
($n = 3.42$ \cite{franta2017temperature}) in a honeycomb lattice (air
background, lattice constant $a = 20~\mu$m, unperturbed rod diameter
$d = 0.46a$), as shown in Fig.~\ref{fig:1}(a--c). The band structure
(Fig.~\ref{fig:1}(e)) exhibits double degeneracy at K and K$'$, forming a
Dirac point with linear dispersion (Dirac cone) in its vicinity, which defines
inequivalent valleys and underpins valley-dependent topological behavior.

\subsection{Inversion Symmetry Breaking and Bandgap Opening}\label{sec:bandgap}

\begin{figure}[H]
    \centering
    \includegraphics[width=1\linewidth]{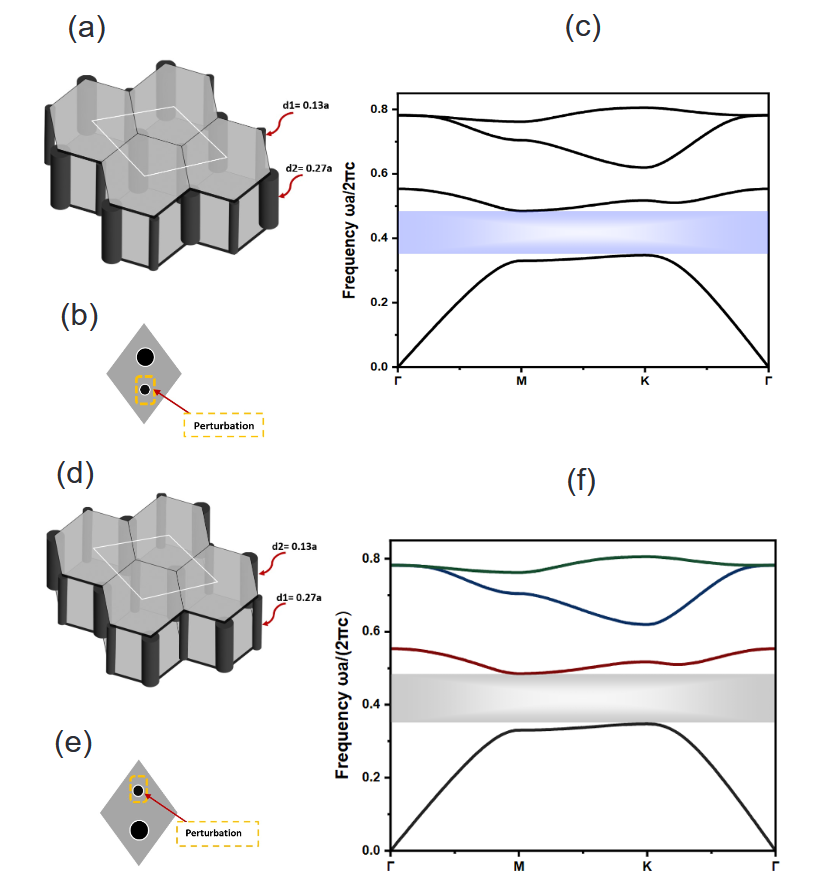}
    \caption{\textbf{Perturbed photonic crystal structures.} (a)~PC-A with
    $d_1 = 0.27a$, $d_2 = 0.13a$. (b)~Rhombic unit cell of PC-A.
    (c)~Band diagram showing bandgap opening. (d)~PC-B with diameters
    exchanged. (e)~Rhombic unit cell of PC-B. (f)~Band diagram of
    PC-B---identical bandgap to PC-A but topologically inverted.}
    \label{fig:2}
\end{figure}

The topological bandgap emerges when alternate rods are assigned diameters
$d_1 = 0.27a$ and $d_2 = 0.13a$, preserving $C_3$ symmetry while explicitly
breaking inversion symmetry. This perturbed structure is designated PC-A
(Fig.~\ref{fig:2}(a--c)). Exchanging the two rod diameters yields the
topologically inequivalent PC-B (Fig.~\ref{fig:2}(d--f)), which shares the
same bandgap width and frequency range as PC-A but carries an inverted Berry
curvature distribution, confirming topological inequivalence.

\subsection{Berry Curvature and Valley Chern Numbers}\label{sec:berry}

The system exhibits non-zero Berry curvature at both K and K$'$ with opposite
signs. Following Zhao et al.\ \cite{zhao2020first}, the Berry curvature for
band $n$ is:
\begin{equation}
    \Omega_n(\mathbf{k}) = \nabla_{\mathbf{k}} \times \mathbf{A}_n(\mathbf{k}),
    \label{eq:berry_curl}
\end{equation}
where $\mathbf{A}_n(\mathbf{k}) = -i\langle u_{n\mathbf{k}} |
\nabla_{\mathbf{k}} | u_{n\mathbf{k}} \rangle$ is the Berry connection for
Bloch mode $|u_{n\mathbf{k}}\rangle$. The normalized Berry curvature
distribution and valley Chern number are presented in Fig.~\ref{fig:1}(d).

The low-energy physics near K/K$'$ is governed by the effective
$\mathbf{k}\cdot\mathbf{p}$ Hamiltonian \cite{ma2016all}:
\begin{equation}
    \hat{H}_{K/K'} = \pm v_D (\sigma_x \delta k_x + \sigma_y \delta k_y)
    \pm \gamma \sigma_z,
    \label{eq:hamiltonian}
\end{equation}
where $v_D$ is the Dirac velocity, $\delta\mathbf{k} = \mathbf{k} - \mathbf{K}$,
$\gamma$ is the symmetry-breaking mass parameter, and $\sigma_{x,y,z}$ are
Pauli matrices on the sublattice pseudospin. Least-squares fitting yields
$\gamma = 0.31$~THz, giving a K-point gap estimate
$\Delta\omega_K = 2|\gamma| = 0.62$~THz. The full photonic bandgap spans
$\omega a/2\pi c \approx 0.36$--$0.49$ ($\approx 5.4$--$7.4$~THz,
width $\approx 1.95$~THz), as visible in Figs.~\ref{fig:2}(c,f). The
$\mathbf{k{\cdot}p}$ Hamiltonian confirms non-trivial topological character
and correct valley Chern numbers; a Wilson-loop approach \cite{yu2011} gives
an equivalent gauge-invariant confirmation.

The valley Chern number integrates Berry curvature over the half-Brillouin
zone (HBZ) \cite{castro2009electronic}:
\begin{equation}
C^{K/K'} = \frac{1}{2\pi} \int_{\mathrm{HBZ}_{K/K'}} \Omega(\mathbf{k})
\, d^2\mathbf{k}.
\label{eq:chern}
\end{equation}
Numerical integration yields $C_K^{\mathrm{PC\text{-}A}} = +\tfrac{1}{2}$,
$C_{K'}^{\mathrm{PC\text{-}A}} = -\tfrac{1}{2}$, with all signs inverted for
PC-B. Time-reversal symmetry enforces $C_K + C_{K'} = 0$; the non-zero
individual values follow from inversion symmetry breaking, confirming PC-A
and PC-B belong to topologically distinct classes.

\subsection{Bulk-Edge Correspondence}\label{sec:bec}

The number of topologically protected edge states at the PC-A/PC-B domain
wall is \cite{ma2016all}:
\begin{equation}
N_{\mathrm{edge}} = \bigl|C_K^{\mathrm{PC\text{-}A}} -
C_K^{\mathrm{PC\text{-}B}}\bigr| =
\bigl|(+\tfrac{1}{2}) - (-\tfrac{1}{2})\bigr| = 1,
\end{equation}
predicting exactly one protected edge state per valley, verified by the single
edge dispersion branch in Fig.~\ref{fig:3}(b). Valley-Hall protection differs
from the absolute protection of Chern insulators: backscattering suppression
relies on K--K$'$ separation in momentum space, so perturbations smooth
relative to the $4\pi/(3a)$ valley distance cannot mix valleys efficiently,
whereas sub-lattice-scale disorder can partially break it \cite{arregui2021}.
The defects here are sub-wavelength yet smooth at the lattice scale, hence
effectively protected; we therefore describe the mechanism as strong suppression rather than absolute prohibition of backscattering.

\section{Simulation Methodology}\label{sec:sim}

All simulations were performed using COMSOL Multiphysics (Wave Optics Module)
with scattering boundary conditions on open boundaries. Band structures are
obtained from eigenfrequency studies with Bloch-periodic boundary conditions;
transmission is computed via frequency-domain port excitation. Q-factors are
extracted from complex eigenfrequencies and verified independently by
Lorentzian fits to the transmission spectra. Sensitivity is evaluated
from resonance wavelength shifts at each analyte refractive index. Berry
curvature and valley Chern numbers are computed in MATLAB following
Zhao et al.\ \cite{zhao2020first}.

\subsection{Mesh Convergence and Numerical Validation}\label{sec:mesh}

Table~\ref{tab:convergence} reports mesh convergence for the hexagonal cavity
(PC-12, $n = 1.395$). The fine mesh ($\lambda/8$ cavity, $\lambda/5$ bulk)
achieves convergence within 0.3\%; $\lambda/10$ refinement changes $Q_L$ by
$<$0.25\%. Q-factors incorporate in-plane radiation loss and silicon material
absorption from the complex refractive index, but exclude out-of-plane losses
(finite rod height, substrate coupling), so reported values are theoretical
figures of merit within the 2D model (Section~\ref{sec:limitations}).
The K-point gap $\Delta\omega_K = 0.62$~THz cross-validates the Berry curvature
methodology; the full photonic bandgap ($\approx 1.95$~THz) encloses all
edge-state and cavity-resonance frequencies.

\begin{table}[htbp]
\caption{Mesh convergence of the cavity loaded $Q_L$ (PC-12 analyte, $n = 1.395$).}
\label{tab:convergence}
\begin{tabular}{@{}lccc@{}}
\toprule
\textbf{Mesh level} & \textbf{Max element size} & $\boldsymbol{Q_L}$\textbf{-factor} & $\boldsymbol{f_0}$ \textbf{(THz)} \\
\midrule
Coarse     & $\lambda/4$  & 241,500 & 5.7821 \\
Medium     & $\lambda/6$  & 278,900 & 5.7834 \\
Fine       & $\lambda/8$  & 285,338 & 5.7841 \\
Extra fine & $\lambda/10$ & 286,012 & 5.7842 \\
\botrule
\end{tabular}
\end{table}

\section{Topological Waveguides}\label{sec:wg}

\begin{figure}[H]
    \centering
    \includegraphics[width=1\linewidth]{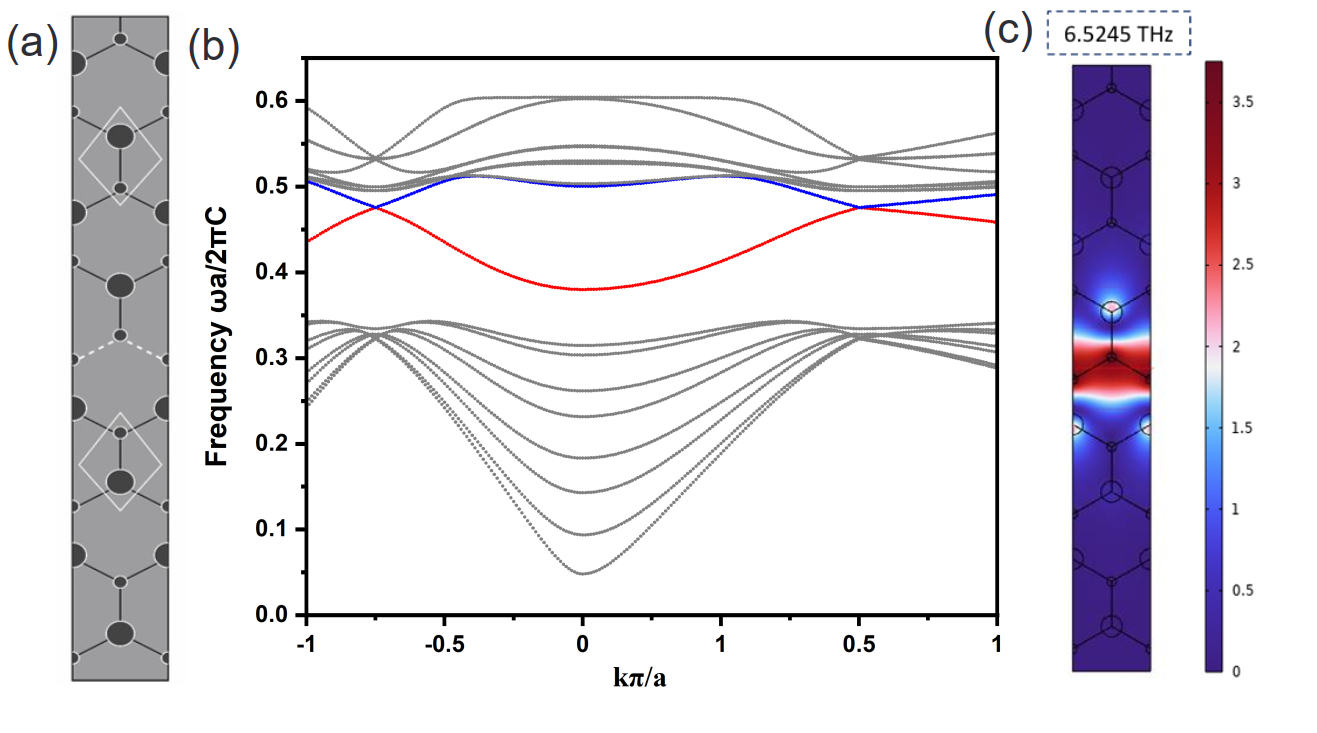}
    \caption{\textbf{Topological edge states at the PC-A/PC-B interface.} (a)
    Domain-wall configuration; interface indicated by white dashed lines. (b)
    Projected band structure with single edge-state branch within the bulk
    bandgap, consistent with $N_{\mathrm{edge}} = 1$. (c) Normalized electric
    field at 6.5245~THz demonstrating interface-localized propagation.}
    \label{fig:3}
\end{figure}

When PC-A and PC-B are placed in contact, valley-polarized edge states emerge
within the shared bulk bandgap. The projected band structure
(Fig.~\ref{fig:3}(b)) contains a single continuous edge-state branch spanning
the bandgap, consistent with $N_{\mathrm{edge}} = 1$, and the electric field
at 6.5245~THz (Fig.~\ref{fig:3}(c)) confirms strong interface localization
with exponential decay into both bulk regions.

\subsection{Linear Waveguide}\label{sec:linear}

\begin{figure}[H]
    \centering
    \includegraphics[width=0.95\linewidth]{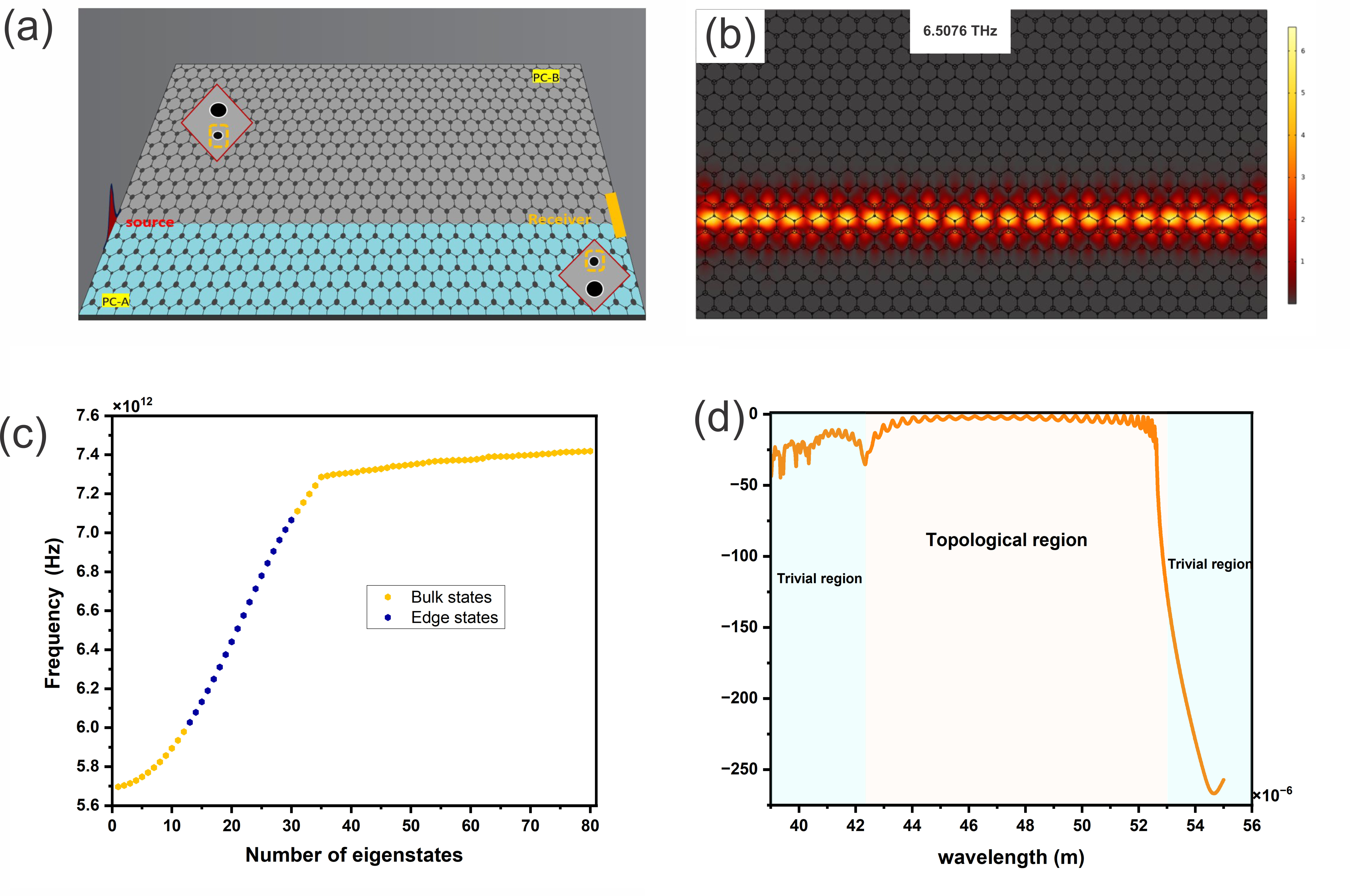}
    \caption{\textbf{Linear waveguide.} (a) Structural schematic. (b) Normalized
    electric field at 6.5076~THz. (c) Frequency vs.\ number of eigenstates.
    (d) Transmission spectrum; the nontrivial bandgap region (shaded) exhibits
    low insertion loss.}
    \label{fig:4}
\end{figure}

A straight waveguide (33$\times$21 unit cells) with the zigzag PC-A/PC-B
interface at its center achieves $-1.12$~dB ($\approx 77$\% power) maximum
transmission in the nontrivial bandgap (Fig.~\ref{fig:4}); trivial bands show
substantially higher loss, confirming valley-polarized transport.

\begin{figure}[H]
    \centering
    \includegraphics[width=0.85\linewidth]{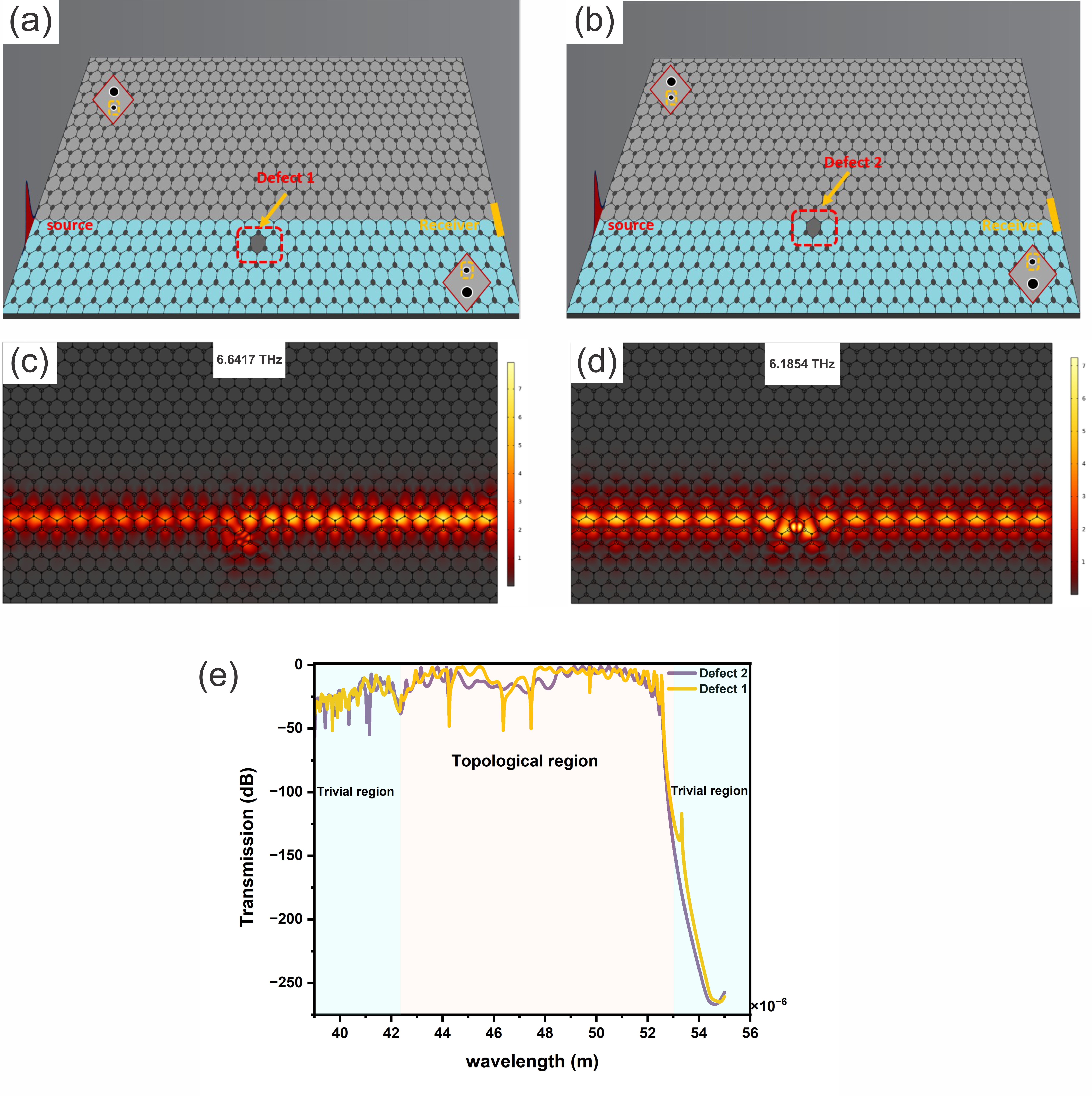}
    \caption{\textbf{Defect tolerance in the linear waveguide.} (a,b) Positions of
    Defects~1 and~2. (c,d) Normalized field distributions confirming
    propagation through defect regions with minor perturbation. (e)
    Transmission spectra showing sustained performance in the nontrivial
    bandgap.}
    \label{fig:5}
\end{figure}

To quantify defect tolerance, two configurations are tested: Defect~1
(further from the domain wall) and Defect~2 (close to the domain wall),
shown in Figs.~\ref{fig:5}(a,b). These represent limiting cases of minimum
and maximum scattering potential on the edge mode; the large bandgap
($\approx 1.95$~THz) theoretically ensures no sub-wavelength perturbation can
close the gap and induce intervalley coupling. Table~\ref{tab:linear_defects}
summarizes the results; worst-case degradation is only 0.47~dB, substantially
below losses reported in comparable non-topological photonic crystal waveguides
under similar disorder \cite{arregui2021,he2019silicon}.

\begin{table}[tbp]
\caption{Straight-waveguide transmission under structural defects.}
\label{tab:linear_defects}
\resizebox{\columnwidth}{!}{%
\begin{tabular}{@{}lccc@{}}
\toprule
\textbf{Configuration} & \textbf{Max transmission (dB)} & \textbf{Power (\%)} & \textbf{Degradation (dB)} \\
\midrule
Pristine  & $-1.12$ & 77.2 & ---     \\
Defect~1  & $-1.14$ & 76.9 & $+0.02$ \\
Defect~2  & $-1.60$ & 69.2 & $+0.47$ \\
\botrule
\end{tabular}
}
\end{table}
\subsection{\texorpdfstring{$\Omega$}{Omega}-Shaped Waveguide}\label{sec:omega}

\begin{figure}[H]
    \centering
    \includegraphics[width=\linewidth]{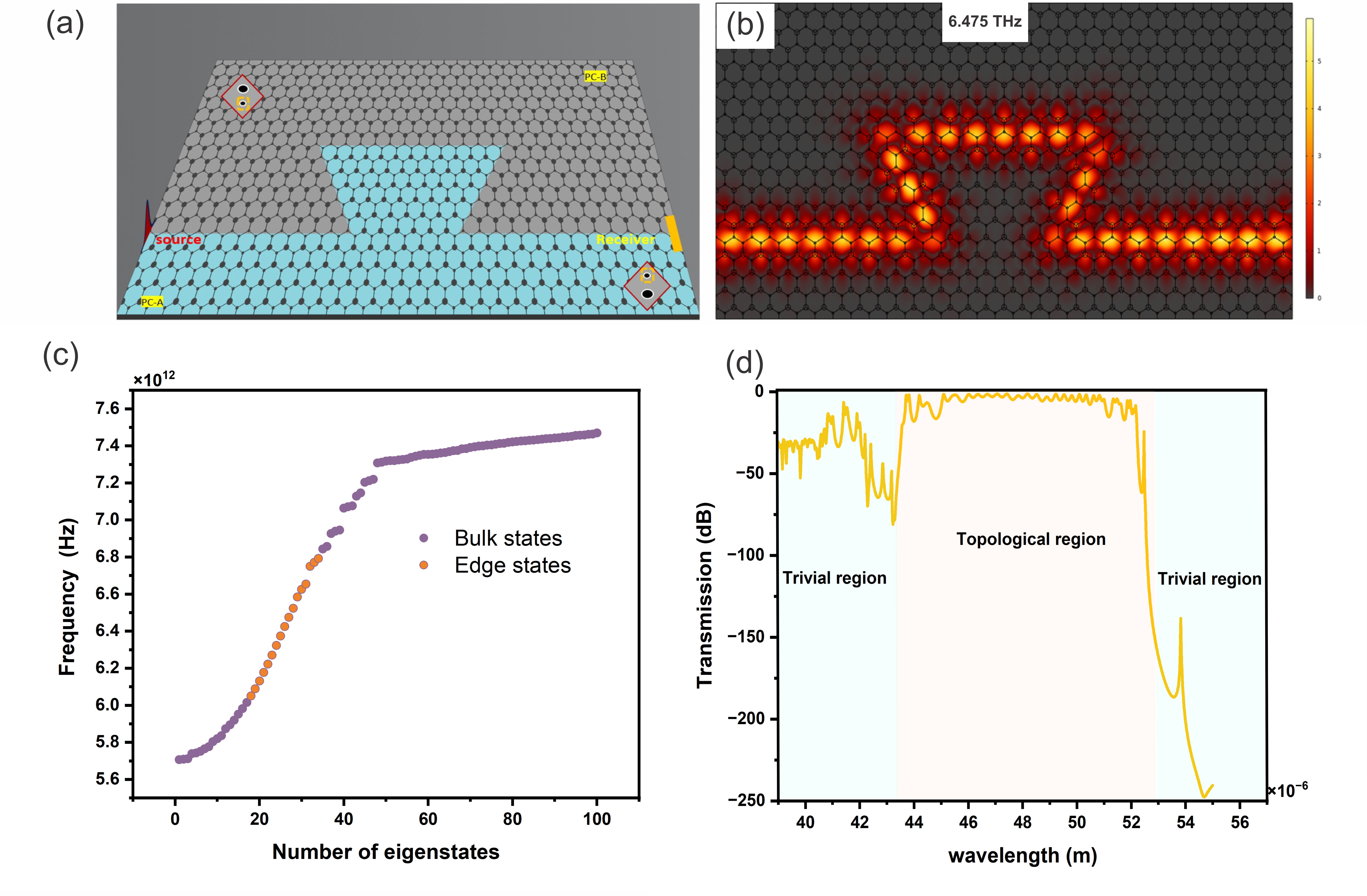}
    \caption{\textbf{$\Omega$-shaped waveguide.} (a) Structural schematic. (b)
    Normalized electric field confirming bend-robust propagation. (c)
    Frequency vs.\ number of eigenstates. (d) Transmission spectrum;
    nontrivial bandgap maintains low insertion loss comparable to the
    straight waveguide.}
    \label{fig:6}
\end{figure}

In order to maintain the zigzag interface, the $\Omega$-shaped waveguide
inserts two 60$^\circ$ bends in PC-A before PC-B integration
(Fig.~\ref{fig:6}(a)). Since sharp bends dominate scattering loss in
conventional waveguides and would necessitate engineered tapers without
topological design, they pose a demanding test of topological protection.
The normalized field (Fig.~\ref{fig:6}(b)) demonstrates that the edge mode
passes through both bends with no discernible back-reflection or radiation.
The two-bend geometry yields a transmission maximum of $-1.27$~dB
($\approx 75$\%, Fig.~\ref{fig:6}(d)), only 0.15~dB below the straight case.

\begin{figure}[H]
    \centering
    \includegraphics[width=\linewidth]{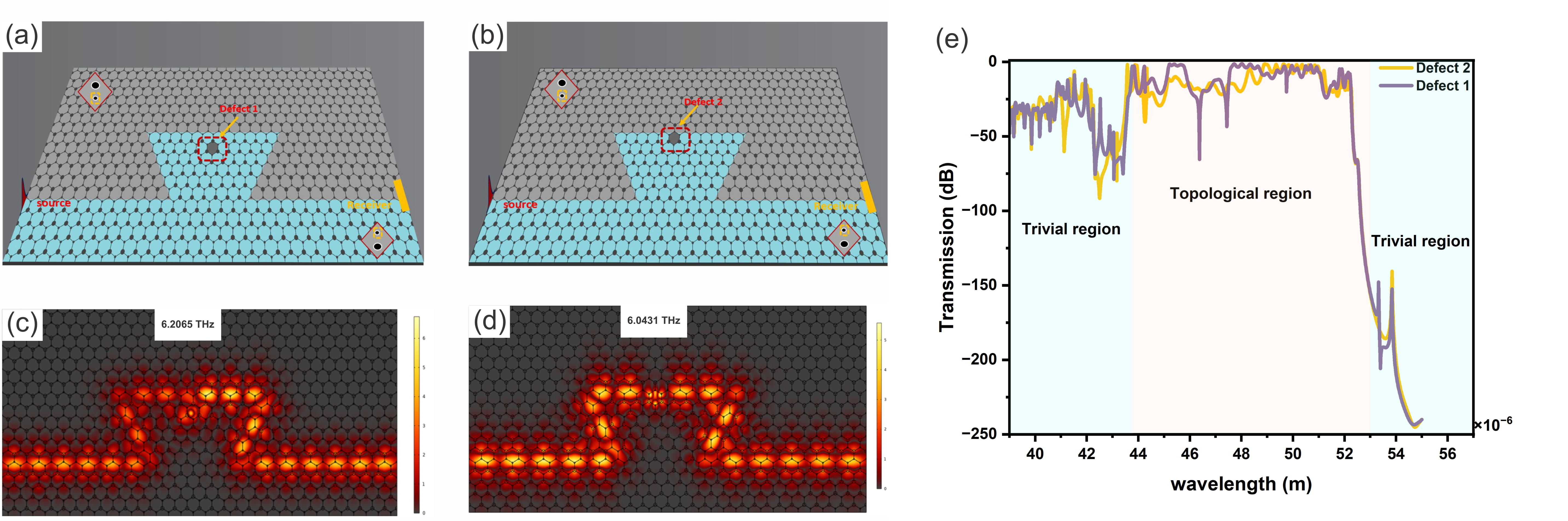}
    \caption{\textbf{Defect tolerance in the $\Omega$-shaped waveguide.} (a,b) Defect
    positions. (c,d) Field distributions confirming propagation through
    simultaneous bends and defects. (e) Transmission spectra demonstrating
    sustained topological protection.}
    \label{fig:7}
\end{figure}

Defects~1 and~2 were also added to the $\Omega$-shaped waveguide to
concurrently stress the edge states with acute bends and localized disorder.
The results are quantified in Table~\ref{tab:omega_defects};
Figs.~\ref{fig:7}(c,d) display the field distributions. Defect-induced
degradation ($<$0.05~dB) is well below the bend penalty (0.15~dB): valley-Hall
protection strongly suppresses backscattering from local disturbances, whereas
geometric scattering from large-scale shape changes is not within the scope of
this topological protection mechanism. All configurations retain transmission
above 73\%.

\begin{table}[tbp]
\caption{$\Omega$-waveguide transmission under structural defects.}
\label{tab:omega_defects}
\resizebox{\columnwidth}{!}{%
\begin{tabular}{@{}lccc@{}}
\toprule
\textbf{Configuration} & \textbf{Max transmission (dB)} & \textbf{Power (\%)} & \textbf{Degradation (dB)} \\
\midrule
Pristine  & $-1.27$ & 74.7 & ---     \\
Defect~1  & $-1.31$ & 73.9 & $+0.04$ \\
Defect~2  & $-1.25$ & 75.0 & $-0.02$ \\
\botrule
\end{tabular}
}
\end{table}
\section{Topological Cavity and Coupled Waveguide Formation}\label{sec:cavity}

\begin{figure}[H]
    \centering
    \includegraphics[width=\linewidth]{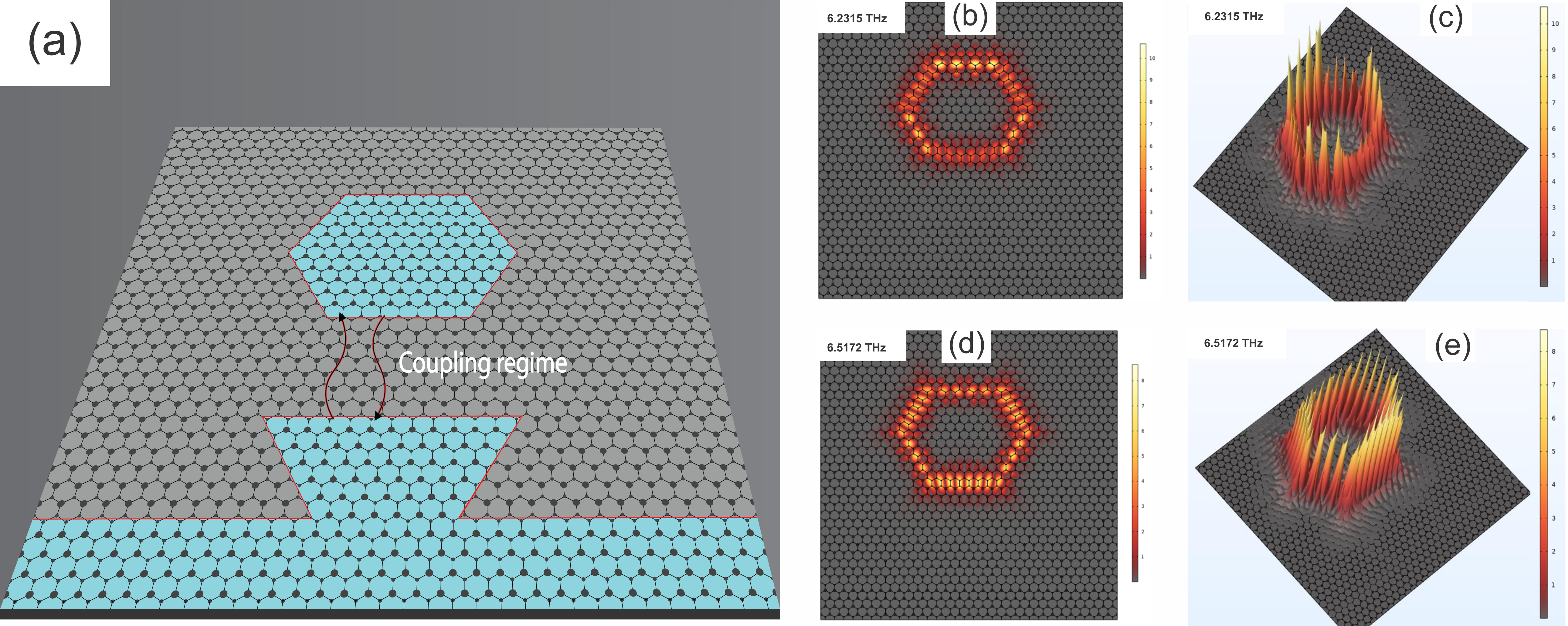}
    \caption{\textbf{Hexagonal topological cavity.} (a) Structural schematic. (b,d)
    Normalized electric field at 6.2315~THz and 6.5172~THz, showing strong
    modal confinement. (c,e) Spatial representation confirming localization.}
    \label{fig:8}
\end{figure}

Embedding PC-A inside PC-B creates a hexagonal closed interface loop
($\approx$6 unit cells per side) supporting whispering-gallery-like resonant
modes. An $\Omega$-shaped waveguide evanescently couples light into the cavity
at resonance; off-resonant frequencies pass through undisturbed. The
topological interface suppresses resonant-mode coupling to bulk radiative
modes, the primary mechanism underlying the high $Q$-factors of
Section~\ref{sec:sensing}. Field distributions at 6.2315 and 6.5172~THz
(Fig.~\ref{fig:8}) confirm strong hexagonal modal confinement with distinct
angular momentum states.

\subsection{Cavity Coupling Analysis}\label{sec:coupling}

The total (loaded) Q-factor $Q_L$ of a resonator side-coupled to a waveguide
is related to the intrinsic Q-factor $Q_i$ and coupling Q-factor $Q_c$ by
\cite{haus1991coupled}:
\begin{equation}
\frac{1}{Q_L} = \frac{1}{Q_i} + \frac{1}{Q_c},
\label{eq:ql}
\end{equation}
where $Q_i$ characterizes intrinsic cavity losses and $Q_c$ characterizes
power leakage into the coupled waveguide. The coupling coefficient
$\kappa = Q_i/Q_c$ is extracted from the on-resonance transmission minimum
\cite{haus1991coupled}:
\begin{equation}
T_{\min} = \left(\frac{\kappa - 1}{\kappa + 1}\right)^2.
\label{eq:tmin}
\end{equation}
The local off-resonance background is estimated by a polynomial fit to the
transmission spectrum on both sides of the resonance dip, and the dip contrast
is measured relative to this baseline. Extracting $\kappa$ via
Eq.~(\ref{eq:tmin}) yields $\kappa = 1.005$--$1.025$ across all five cell
lines (Table~\ref{tab:coupling}). The near-unity $\kappa > 1$ confirms slight
over-coupling throughout the analyte range, consistent with the 38--52~dB
transmission dip contrasts observed at resonance and ensuring efficient signal
readout while maintaining high $Q_L$.

\begin{table}[tbp]
\caption{Coupling parameters for the five cancer cell-line analytes.}
\label{tab:coupling}
\resizebox{\columnwidth}{!}{%
\begin{tabular}{@{}lccccc@{}}
\toprule
\textbf{Cell} & $Q_L$ & $Q_i$ & $Q_c$ & $\kappa$ & \textbf{Contrast (dB)} \\
\midrule
Jurkat     & 141,643 & 286,807 & 279,850 & 1.025 & 38.2 \\
HeLa       &  90,977 & 182,404 & 181,505 & 1.005 & 52.1 \\
PC-12      & 285,338 & 573,003 & 568,367 & 1.008 & 47.8 \\
MDA-MB-231 & 159,025 & 320,714 & 315,429 & 1.017 & 41.6 \\
MCF-7      & 199,173 & 400,333 & 396,377 & 1.010 & 46.1 \\
\botrule
\end{tabular}
}
\end{table}

\subsection{Modal Volume and Cavity Figure of Merit}\label{sec:modalvol}

To characterize light confinement beyond the Q-factor alone, we evaluate
the effective modal volume \cite{joannopoulos2008molding}:
\begin{equation}
    V_m = \frac{\displaystyle\int \epsilon(\mathbf{r})\,
    |\mathbf{E}(\mathbf{r})|^2\,dV}
    {\displaystyle\max\bigl[\epsilon(\mathbf{r})\,
    |\mathbf{E}(\mathbf{r})|^2\bigr]},
    \label{eq:vm}
\end{equation}
where $\epsilon(\mathbf{r})$ is the local permittivity and
$\mathbf{E}(\mathbf{r})$ is the normalized electric field distribution.
Since the simulations are 2D, the volume integrals reduce to area integrals;
$V_m$ thus carries an implicit unit-length factor along the rod axis and is
expressed in units of $(\lambda_0)^3$, as is standard for 2D PC cavities
\cite{akahane2003high}. Numerator and denominator come directly from the
field solution at the PC-12 resonance by
evaluating the surface-integral and maximum operators over the simulation
domain. Evaluating Eq.~(\ref{eq:vm}) yields:
\begin{equation}
    V_m = 2.27\,\left(\frac{\lambda_0}{n_{\rm eff}}\right)^3,
\end{equation}
where $n_{\rm eff} = \sqrt{\bar{\epsilon}}$ is the area-weighted effective
permittivity of the PC-A region. With silicon fill fraction
$f_{\rm Si} = 8.14\%$ (from rod diameters $d_1 = 0.27a$, $d_2 = 0.13a$ in
the rhombic unit cell of area $\sqrt{3}a^2/2$), the effective permittivity is
$\bar{\epsilon} = f_{\rm Si}\,n_{\rm Si}^2 + (1-f_{\rm Si})\cdot 1 = 1.871$,
giving $n_{\rm eff} = 1.37$. The cavity figure of merit is:
\begin{equation}
    \frac{Q}{V_m} = \frac{285{,}338}{2.27\,(\lambda_0/1.37)^3}
    = 1.26\times10^5 \;\left(\frac{\lambda_0}{n_{\rm eff}}\right)^{-3}.
    \label{eq:qv}
\end{equation}
This large $Q/V_m$ ratio indicates strong electromagnetic energy confinement
relative to the mode volume, directly enhancing light--matter interaction in
the analyte-occupied air regions and contributing to the high FoM reported
in Section~\ref{sec:sensing}.

\section{Sensing}\label{sec:sensing}

\subsection{Sensing Principle and Perturbation Theory Consistency
}\label{sec:principle}

The sensor operates on resonance wavelength tracking: an analyte shifts the
resonant frequency proportionally to $\Delta n$. Topology contributes by
(i) suppressing radiative loss at the cavity boundary to sustain high $Q_L$,
and (ii) keeping $Q_L$ stable under fabrication disorder, as demonstrated by
the defect-tolerance results of Section~\ref{sec:linear}. The volumetric field
confinement factor is:
\begin{equation}
\Gamma = \frac{\int_{\mathrm{analyte}} |\mathbf{E}|^2 \, dV}
{\int_{\mathrm{total}} |\mathbf{E}|^2 \, dV}.
\label{eq:gamma}
\end{equation}
Evaluating directly from the field solution at the PC-12 resonance
($\lambda_0 = 51.963~\mu$m, i.e.\ $\lambda_0 = 51{,}963$~nm):
\begin{equation}
    \Gamma = \frac{19.88}{29.09} = 0.68,
\end{equation}
corresponding to 74\% of the geometric upper bound
($\Gamma_{\rm max} = 1 - f_{\rm Si} = 0.919$, where $f_{\rm Si} \approx 8.1\%$).
$\Gamma = 0.68$ represents a substantially higher confinement factor than
surface-evanescent designs ($\Gamma < 0.20$) \cite{kumar2022topological}.

\textbf{Perturbation theory consistency .}
As an analytic consistency check, we apply first-order electromagnetic
perturbation theory \cite{joannopoulos2008molding}. For a small permittivity
perturbation $\Delta\epsilon$ confined to the analyte volume, the fractional
resonance shift is:
\begin{equation}
    \frac{\Delta\lambda}{\lambda_0} = +\frac{1}{2}
    \frac{\displaystyle\int_{V_a} \Delta\epsilon(\mathbf{r})\,
    |\mathbf{E}(\mathbf{r})|^2\,dV}
    {\displaystyle\int_{\rm total} \epsilon(\mathbf{r})\,
    |\mathbf{E}(\mathbf{r})|^2\,dV},
    \label{eq:pt}
\end{equation}
where $\Delta\epsilon = n_{\rm analyte}^2 - n_{\rm air}^2$ and the
unperturbed field $\mathbf{E}$ is taken from the full-wave solution with
air background ($n = 1$). Differentiating Eq.~(\ref{eq:pt}) with respect
to $n$ and using $\lambda_0 = 51{,}963$~nm:
\begin{equation}
    S_{\rm PT} = \lambda_0 \cdot
    \frac{n_{\rm analyte}\,W_{\rm analyte}}
    {n_{\rm Si}^2\,W_{\rm Si} + n_{\rm analyte}^2\,W_{\rm analyte}}
    \approx 9{,}840~\text{nm/RIU},
    \label{eq:s_pt}
\end{equation}
where $W_{\rm analyte} = 19.88$ and $W_{\rm Si} = 9.21$ are the
extracted $\int|\mathbf{E}|^2\,dV$ contributions from analyte and
silicon regions respectively. The first-order PT estimate is a
consistency check on the FEM sensitivity (24,171~nm/RIU): it systematically
underestimates the FEM value because it uses the unperturbed air-background
field and cannot capture the redistribution that occurs when the cavity is
loaded with a finite-permittivity analyte. As permittivity increases from
unity, the field shifts toward the analyte \cite{joannopoulos2008molding},
raising $\Gamma$ and the sensitivity above the PT estimate. The ratio
$S_{\rm FEM}/S_{\rm PT} \approx 2.5$ is consistent with this redistribution for
the large $\Delta n \approx 0.4$ loading ($n_{\rm analyte} \approx 1.395$),
which lies beyond the first-order regime; the FEM value is thus authoritative
and the PT estimate serves only as a self-consistency check.

The RI values used here are experimentally reported bulk values
\cite{liang2007determining}; a full implementation would require biological
medium modelling, fluidic integration, and cell-morphology characterisation,
and must contend with strong THz water absorption at 6~THz
(Section~\ref{sec:limitations}). The hexagonal cavity is compatible with
microfluidic integration, and the large air fraction (91.9\%) within PC-A
provides ample void volume for analyte infiltration.

\subsection{Biosensing Results}\label{sec:biosensing}

The five cell lines serve as representative benchmark analytes
with closely spaced refractive indices (1.390--1.401~RIU,
Table~\ref{tab:cell_refractive_indices}) \cite{liang2007determining}; they are
of heterogeneous origin (leukemia, carcinomas, and a pheochromocytoma) and
are used here purely as fixed-RI test points.

\begin{figure}[H]
    \centering
    \includegraphics[width=\linewidth]{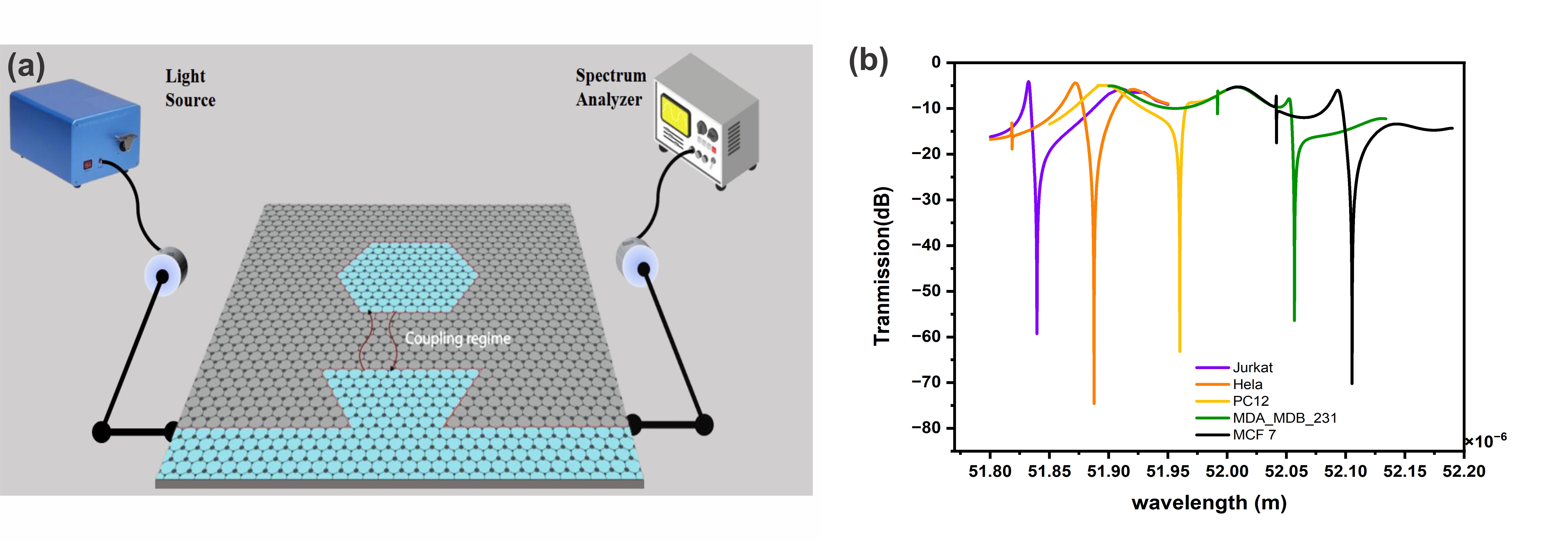}
    \caption{\textbf{Cancer cell-line RI discrimination.} (a) Proposed experimental
    setup. (b) Transmission spectra for five cancer cell lines demonstrating
    systematic monotonic resonance shifts with refractive index.}
    \label{fig:9}
\end{figure}

The five cancer cell lines produce monotonic resonance shifts with increasing
refractive index (Fig.~\ref{fig:9}). Q-factor and sensitivity are:
\begin{equation}
Q_L = \frac{\lambda_o}{\mathrm{FWHM}}, \qquad S = \frac{\Delta\lambda}{\Delta n}.
\end{equation}

\begin{table}[htbp]
\caption{Refractive indices of the cancer cell lines (80\% density) \cite{liang2007determining}.}
\label{tab:cell_refractive_indices}
\begin{tabular}{@{}llc@{}}
\toprule
\textbf{Cell line}  & \textbf{Tumor origin}       & \textbf{RI} \\
\midrule
Jurkat     & T-cell leukemia         & 1.390 \\
HeLa       & Cervical adenocarcinoma & 1.392 \\
PC-12      & Pheochromocytoma        & 1.395 \\
MDA-MB-231 & Breast adenocarcinoma   & 1.399 \\
MCF-7      & Breast adenocarcinoma   & 1.401 \\
\botrule
\end{tabular}
\end{table}

Table~\ref{tab:cell_data} presents resonance wavelengths, $Q_L$-factors, FWHM,
and FoM $= \bar{S}/\mathrm{FWHM}$ for the five cell lines. Resonant wavelengths
span 51.84--52.11~$\mu$m (51,840--52,110~nm). $Q_L$ reaches $285{,}338$ for
PC-12; HeLa has the broadest linewidth (0.570~nm) and
lowest $Q_L$ (90,977). Maximum pairwise sensitivity is 24,300~nm/RIU
(MDA-MB-231$\to$MCF-7), average 24,171~nm/RIU. The high absolute sensitivity
follows from the long operating wavelength ($\lambda_0 \approx 52~\mu$m
$= 52{,}000$~nm). FoM is computed with $\bar{S} = 24{,}171$~nm/RIU and FWHM
in nm; maximum FoM of 132,813~RIU$^{-1}$ is attained by PC-12 (narrowest
linewidth). $Q_L$ non-monotonicity is discussed in Section~\ref{sec:qnonmono}.

\begin{table}[tbp]
\caption{Resonance characteristics, $a = 20~\mu$m design; FoM $= \bar{S}/\mathrm{FWHM}$, $\bar{S} = 24{,}171$~nm/RIU.}
\label{tab:cell_data}
\resizebox{\columnwidth}{!}{%
\begin{tabular}{@{}lcccc@{}}
\toprule
\textbf{Cell} & $\lambda_o$ ($\mu$m) & $\boldsymbol{Q_L}$\textbf{-factor} & \textbf{FWHM (nm)} & \textbf{FoM (RIU$^{-1}$)} \\
\midrule
Jurkat     & 51.8395 & 141,643 & 0.366 &  66,042 \\
HeLa       & 51.8876 &  90,977 & 0.570 &  42,405 \\
PC-12      & 51.9600 & 285,338 & 0.182 & 132,813 \\
MDA-MB-231 & 52.0568 & 159,025 & 0.327 &  73,917 \\
MCF-7      & 52.1054 & 199,173 & 0.262 &  92,218 \\
\botrule
\end{tabular}
}
\end{table}

\begin{figure}[H]
    \centering
    \includegraphics[width=\linewidth]{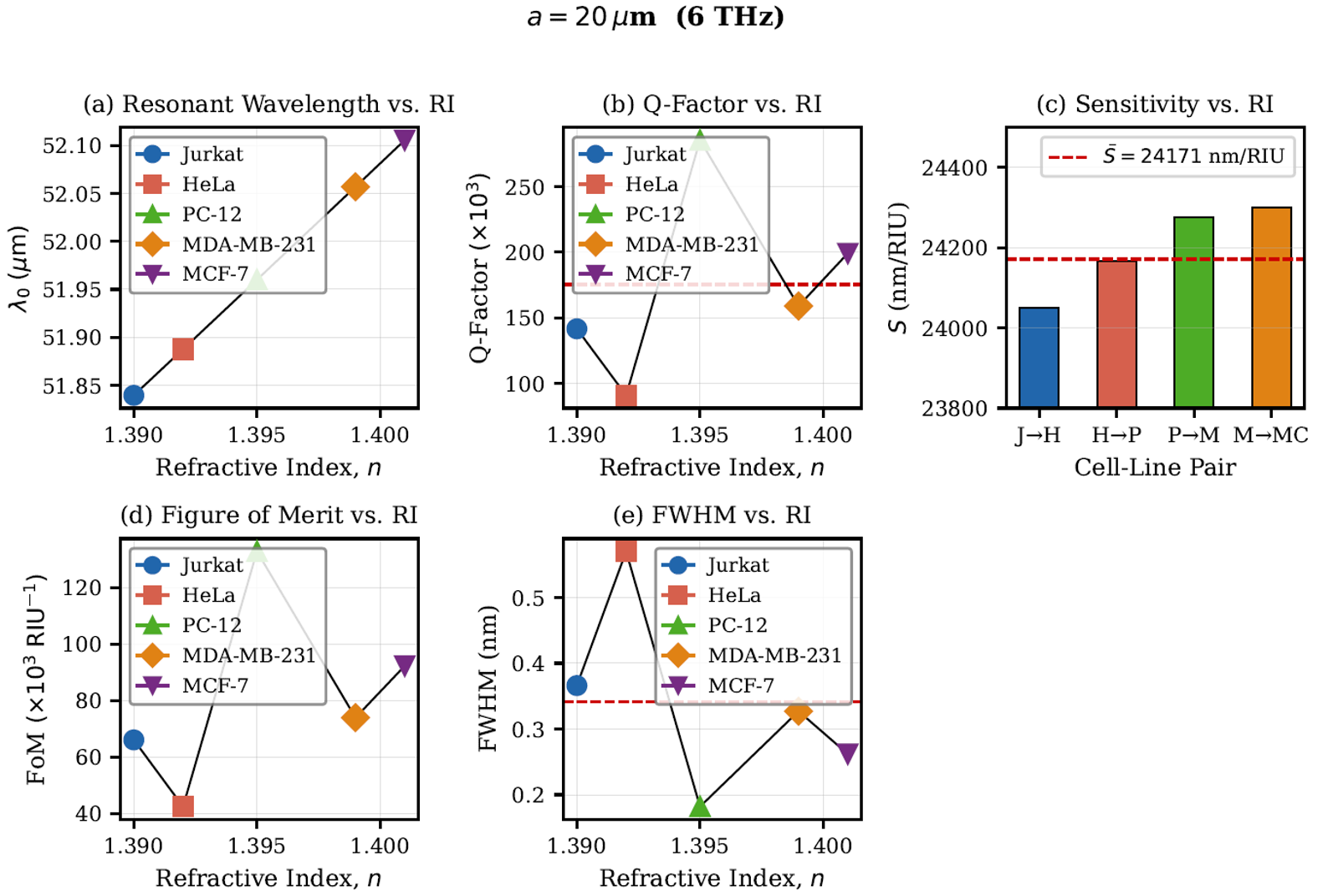}
    \caption{\textbf{Summary of sensing performance metrics for the $a = 20~\mu$m
    design.} (a) Resonant wavelength $\lambda_0$ showing monotonic red-shift.
    (b) $Q_L$-factor, maximal for PC-12 and minimal for HeLa. (c) Pairwise sensitivity $S$ with average
    $\bar{S} = 24{,}171$~nm/RIU (dashed). (d) FoM mirroring $Q_L$-factor trend.
    (e) FWHM showing inverse trend to $Q_L$.}
    \label{fig:10}
\end{figure}

Figure~\ref{fig:10} summarizes five sensing performance metrics as a
function of analyte refractive index for the $a = 20~\mu$m design.

\subsection{Discrimination Margin and Error Analysis}\label{sec:lod}

Using the refractive indices reported by Liang et al.\
\cite{liang2007determining}, the five resonance dips are spectrally
well separated: the inter-cell shifts (48--97~nm, Table~\ref{tab:cell_data})
exceed the resonance linewidths (FWHM $= 0.18$--$0.57$~nm) by approximately two orders
of magnitude, giving non-overlapping dips. This separation is a property of the
fixed input indices used here and should not be read as an achievable
biological resolving power: the 0.25\% uncertainty on the source values
propagates to a wavelength uncertainty of $\approx 84$~nm per
resonance---comparable to, and for the closest pair (48~nm) larger than, the
smallest inter-cell spacing. The result is therefore best interpreted as
high-resolution refractive-index sensing rather than definitive biological
classification; practical cell-line identification would additionally require
THz-specific RI and absorption data and sample-preparation validation.

\section{Discussion}\label{sec:discussion}

\subsection{Q-Factor Non-Monotonicity}\label{sec:qnonmono}

The loaded quality factor $Q_L$ varies non-monotonically with refractive
index: HeLa ($n = 1.392$) yields the lowest value (90,977) and PC-12
($n = 1.395$) the highest (285,338), a threefold spread over $\Delta n = 0.003$. All five resonances lie close together near 5.75--5.78~THz,
well within the 5.4--7.4~THz bandgap, so the variation does not follow a simple
band-edge-proximity trend. It is instead attributed to analyte-dependent changes
in modal symmetry, radiative leakage, and cavity--waveguide coupling within the
gap \cite{barczyk2022interplay}, rather than a monotonic refractive-index
effect. Consistent with this, $\kappa$ stays within 1.005--1.025
(Table~\ref{tab:coupling}), so near-critical coupling is preserved independently
of the resonance shift.

\subsection{Role of Topology in Sensing Performance}\label{sec:topo_role}

The resonance-shift mechanism is common to all high-$Q$ RI sensors and is not
uniquely topological. Topology contributes in two measurable ways: (i) the
interface suppresses radiative loss, enabling $Q_L = 2.85\times10^5$, which lies above
the prior topological THz benchmark \cite{navaratna2023chip} by $>$42\% and at
the upper end of conventional PC defect-cavity ranges ($10^3$--$10^5$
\cite{akahane2003high}); and (ii) fabrication disorder degrades transmission by
at most 0.47~dB (worst-case Defect~2, straight waveguide), stabilising $Q_L$,
FWHM, and $S$ under realistic imperfections.

\subsection{Selectivity and Interference Considerations}\label{sec:selectivity}

The sensor discriminates by refractive index alone, so other substances with
overlapping RI could produce competing shifts. Two factors mitigate this: the
sub-linewidth resolving power (FWHM $<0.6$~nm versus inter-cell shifts
$\geq 48$~nm) gives large margin against background fluctuations, and
microfluidic size-selective filtration or centrifugation can enrich the target
population beforehand. Distinguishing cancerous from non-cancerous cells of the
same tissue would additionally require THz-specific RI data for those pairs
or functionalization such as antibody-coated rod surfaces.

\subsection{Effect of Lattice Constant on Sensing
Performance}\label{sec:scaling}

The sensitivity of a resonance-shift biosensor scales with the operating
wavelength as $S \propto \lambda_0$, a consequence of the proportional scaling
of cavity modal volume with lattice constant $a$. To verify this scaling
behavior and demonstrate design flexibility, we simulated the same topological
cavity architecture with $a = 1~\mu$m, shifting operation to
$\lambda_0 \approx 3310$~nm ($\sim$90~THz) with the same five cancer cell
lines \cite{liang2007determining}.

Table~\ref{tab:scaling} compares the two designs quantitatively. The
$a = 1~\mu$m design yields average sensitivity 885~nm/RIU and
maximum $Q_L = 41{,}354$, approximately 27$\times$ and 6.9$\times$ lower than
the $a = 20~\mu$m design respectively. The sensitivity reduction (27$\times$)
is broadly consistent with the shorter operating wavelength
($\lambda_0$ ratio 15.7$\times$), but exceeds the pure $S \propto \lambda_0$
expectation, indicating that modal overlap, cavity confinement, and
analyte-filling efficiency also change with lattice constant; the more modest
$Q_L$ reduction similarly reflects a weaker-than-linear dependence of radiation
loss on cavity size. Near-critical over-coupling ($\kappa \approx 1.0$) is
preserved across both designs and all analytes (Tables~\ref{tab:coupling}
and~\ref{tab:scaling_coupling}), confirming it is a robust intrinsic property
of the VPhC cavity architecture. These results confirm that
$a = 20~\mu$m remains the preferable lattice constant for cancer cell-line
discrimination, combining both higher sensitivity and higher $Q_L$, whereas
smaller lattice constants may be better suited for molecular or thin-film
analytes at near-infrared wavelengths.

\begin{table}[tbp]
\caption{Performance comparison of the $a = 20$ and $1~\mu$m designs.}
\label{tab:scaling}
\resizebox{\columnwidth}{!}{%
\begin{tabular}{@{}lcccp{3.8cm}@{}}
\toprule
\textbf{Parameter} & $a = 20~\mu$\textbf{m} & $a = 1~\mu$\textbf{m} & \textbf{Ratio} & \textbf{Physical reason} \\
\midrule
$\lambda_0$   & $\sim$52~$\mu$m   & $\sim$3.31~$\mu$m & 15.7$\times$ & Linear scaling with $a$ \\
Max $Q_L$     & 285,338           & 41,354            & 6.9$\times$  & Larger cavity $\Rightarrow$ lower radiation loss \\
$S$ (avg.)    & 24,171~nm/RIU     & 885~nm/RIU        & 27$\times$   & $S \propto \lambda_0$, reduced modal overlap \\
$\kappa$      & 1.005--1.025      & 1.001--1.035      & $\sim$1      & Near-critical coupling preserved \\
Cavity size   & $\sim$120~$\mu$m  & $\sim$6~$\mu$m    & 20$\times$   & Scales with $a$ \\
Cell fit      & Full ($\checkmark$) & Partial ($\times$) & ---        & Cell dia.\ $\sim$10~$\mu$m \\
\botrule
\end{tabular}
}
\end{table}

\begin{table}[tbp]
\caption{Resonance and coupling parameters, $a = 1~\mu$m design ($\bar{S} = 885$~nm/RIU).}
\label{tab:scaling_coupling}
\resizebox{\columnwidth}{!}{%
\begin{tabular}{@{}lcccccc@{}}
\toprule
\textbf{Cell} & $n$ & $\lambda_{\rm res}$ (nm) & $Q_L$ & $Q_i$ & $Q_c$ & $\kappa$ \\
\midrule
Jurkat     & 1.390 & 3306.4 & 34,478 & 69,094 & 68,819 & 1.004 \\
HeLa       & 1.392 & 3310.1 & 20,008 & 40,656 & 39,396 & 1.032 \\
PC-12      & 1.395 & 3310.8 & 22,285 & 45,083 & 44,069 & 1.023 \\
MDA-MB-231 & 1.399 & 3312.0 & 18,969 & 38,602 & 37,297 & 1.035 \\
MCF-7      & 1.401 & 3314.3 & 41,354 & 82,749 & 82,667 & 1.001 \\
\botrule
\end{tabular}
}
\end{table}

\begin{figure}[H]
    \centering
    \includegraphics[width=0.8\linewidth]{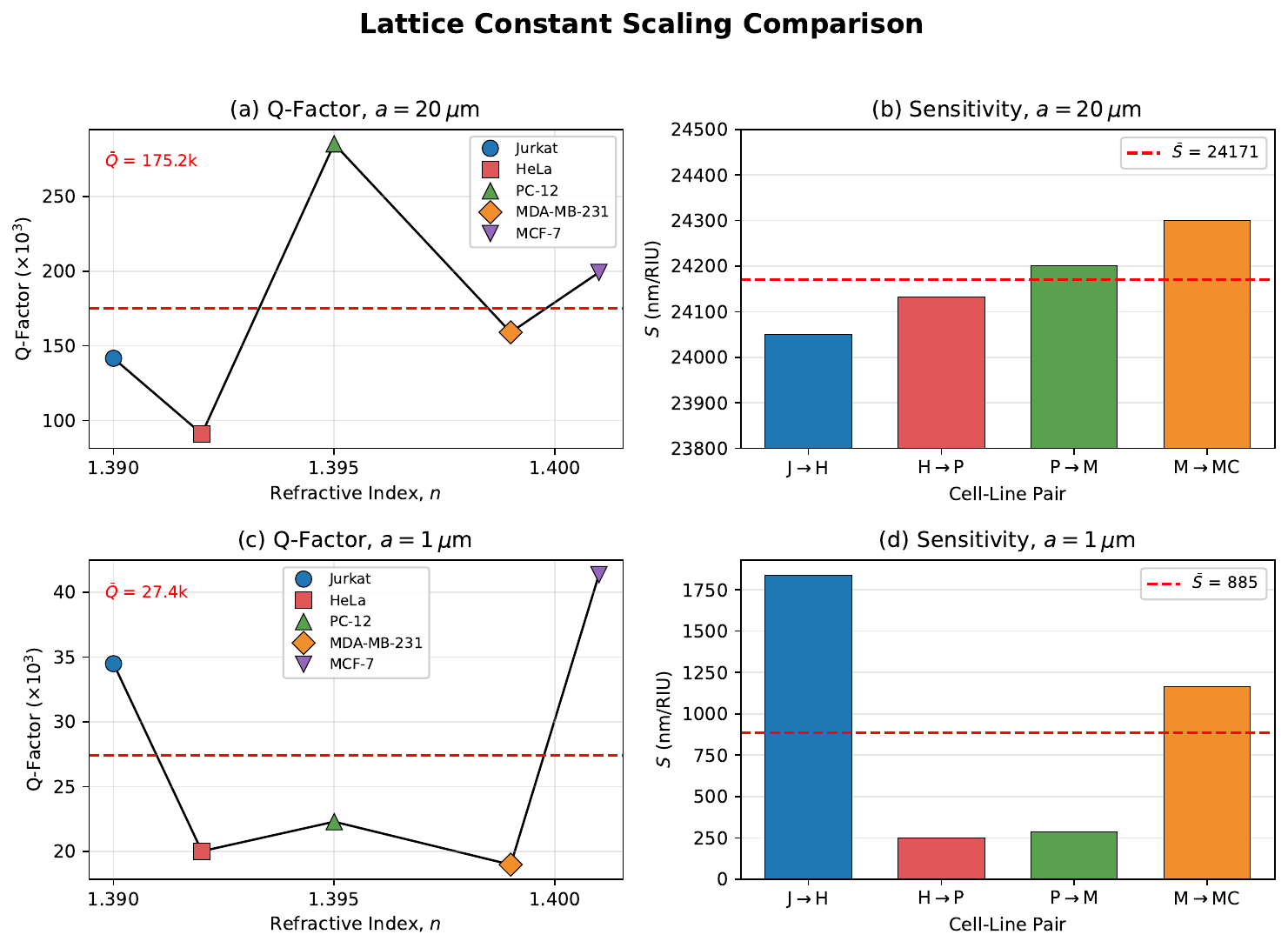}
    \caption{\textbf{Lattice constant scaling comparison between $a = 20~\mu$m
    (top row, panels a--b) and $a = 1~\mu$m (bottom row, panels c--d).} (a,c) $Q_L$-factor
    versus refractive index; (b,d) pairwise sensitivity per cell-line pair.}
    \label{fig:11}
\end{figure}

Figure~\ref{fig:11} presents the scaling comparison.

\subsection{Performance Comparison}\label{sec:comparison}\label{sec:limitations}

Table~\ref{tab:sensor_comparison} benchmarks the proposed design against
representative topological photonic-crystal refractive-index and biosensors
spanning THz, mid-infrared, and telecom bands. Magnetic-field sensors and
hollow-core fiber sensors are excluded as they operate on fundamentally
different physical mechanisms and do not constitute valid performance
benchmarks for refractive-index sensing \cite{fan2008sensitive}. The VPhC
cavity achieves the highest $Q_L$ ($2.85\times10^5$) among the representative
topological biosensors compared in Table~\ref{tab:sensor_comparison},
exceeding the compared THz benchmark \cite{navaratna2023chip}
by more than 42\% and the valley-photonic-crystal ring-resonator biosensor of
Satyaraj et al.\ \cite{satyaraj2024} ($Q = 2.9\times10^4$), all within the
present 2D simulation framework, with volumetric loading ($\Gamma = 0.68$,
substantially higher than evanescent designs \cite{kumar2022topological}).
Sensitivity of
24,300~nm/RIU exceeds the THz metamaterial benchmark \cite{bhati2022ultra}
by 87\% and the recent graded-index topological resonators of
Dash et al.\ \cite{dash2025} and Barvestani \cite{barvestani2025}; a portion
of this margin reflects the longer operating wavelength, so the $Q_L$-factor
and confinement factor---which are wavelength-independent figures of
merit---provide the more decisive advantage over the topological biosensors
compared here.
\begin{table*}[tbp]
\caption{Comparison with representative topological RI/biosensors. N.R.\ = not
reported. Absolute wavelength sensitivity scales with the operating wavelength,
so the $Q_L$ and FoM columns are the wavelength-independent comparators.
FoM $=\bar{S}/\mathrm{FWHM}$; ``This work'' FoM is the PC-12 maximum.}
\label{tab:sensor_comparison}
\small
\begin{tabular}{@{}llccccc@{}}
\toprule
\textbf{Sensor type} & \textbf{Analyte} & \textbf{Band} &
\textbf{Sensitivity} & $\boldsymbol{Q_L}$ &
\textbf{FoM (RIU$^{-1}$)} & \textbf{Ref.} \\
\midrule
THz metamaterial        & RI film            & THz              & 13{,}000~nm/RIU          & N.R.            & N.R.             & \cite{bhati2022ultra} \\
Topo.\ Si on-chip       & Polyimide film     & THz              & N.R.                     & $1.4\times10^5$ & N.R.             & \cite{kumar2022topological} \\
Topo.\ THz cavity       & Solvent/hydration  & 0.3--0.6~THz     & N.R.                     & $2.0\times10^5$ & N.R.             & \cite{navaratna2023chip} \\
Topo.\ VPhC ring        & Brain-tumor cells  & Mid-IR           & 9021~nm/RIU             & $2.9\times10^4$ & N.R.             & \cite{satyaraj2024} \\
Graded topo.\ resonator & Cancer cells       & $\sim$1.5~$\mu$m & 1806~nm/RIU             & $4.4\times10^3$ & 4030             & \cite{dash2025} \\
1D Zak-phase interface  & RI analyte         & Near-IR          & N.R.                     & N.R.            & $2.94\times10^4$ & \cite{anjineya2025} \\
1D topo.\ interface     & Aqueous / glucose  & Near-IR          & 745~nm/RIU              & N.R.            & N.R.             & \cite{barvestani2025} \\
\textbf{This work}      & Cancer cells       & $\sim$6~THz      & \textbf{24{,}300~nm/RIU} & $\mathbf{2.85\times10^5}$ & \textbf{132{,}813} & --- \\
\botrule
\end{tabular}
\end{table*}


\textit{Single effective RI.} Modelling each optically inhomogeneous cell line
by one bulk RI \cite{liang2007determining} is a standard first-order
approximation in computational biosensing
\cite{navaratna2023chip,kumar2022topological}.

\textit{THz water absorption.} Liquid water absorbs strongly at 6~THz
($\alpha \approx 200$--$500$~cm$^{-1}$ \cite{pickwell2006biomedical}), so
aqueous operation would require dried/lyophilized samples, path-length
confinement, or controlled-humidity flow \cite{kumar2022topological}. The RI
values used are optical-frequency measurements; THz-frequency RI and absorption
of each line need independent characterisation. The study should thus be read
as a platform demonstration, with the cell lines as narrow-RI-window benchmark
analytes rather than validated THz targets.
\vspace{-2mm}
\subsection{Fabrication Feasibility}\label{sec:fabrication}
\begin{figure}[H]
    \centering
    \includegraphics[width=\linewidth]{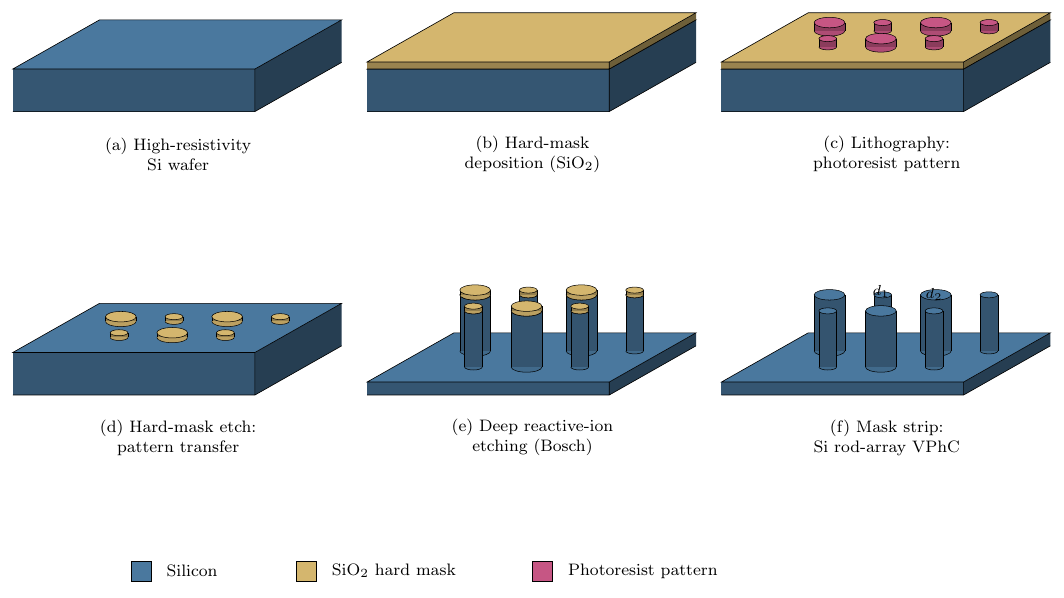}
    \caption{\textbf{Schematic fabrication flow, not to scale, of the silicon
    rod-array valley-Hall photonic crystal:} (a) high-resistivity silicon wafer,
    (b) SiO$_2$ hard-mask deposition, (c) lithographic definition of the
    honeycomb pattern, (d) pattern transfer into the hard mask, (e) Bosch
    deep reactive-ion etching of silicon, and (f) hard-mask removal to obtain
    the silicon rod-array VPhC.}
    \label{fig:fabrication}
\end{figure}

Figure~\ref{fig:fabrication} outlines a representative fabrication flow for the
silicon rod-array VPhC: a high-resistivity Si wafer, SiO$_2$ hard-mask
deposition, lithographic definition of the honeycomb pattern, pattern transfer
into the mask, Bosch deep reactive-ion etching of the rods, and hard-mask
removal. The flow is schematic; etch depth, sidewall roughness, rod-height
uniformity, and substrate release are left for future experimental work.

\section{Conclusion}\label{sec:conclusion}

We proposed a honeycomb valley-Hall photonic crystal cavity for
proof-of-concept refractive-index discrimination of cancer-derived cell lines
at THz frequencies. Topological protection---from valley Chern numbers
$C_v = \pm\tfrac{1}{2}$ with $N_{\mathrm{edge}} = 1$---suppresses intervalley
backscattering in linear and $\Omega$-shaped waveguides through dual
60$^\circ$ bends and structural defects, with worst-case degradation of only
0.47~dB. The hexagonal cavity, with volumetric loading ($\Gamma = 0.68$
versus $<$0.20 for evanescent designs), reaches a
theoretical 2D $Q_L = 285{,}338$, $V_m = 2.27\,(\lambda/n_{\rm eff})^3$, and
$Q/V_m = 1.26\times10^5\,(\lambda/n_{\rm eff})^{-3}$ under near-critical
over-coupling ($\kappa = 1.005$--$1.025$). First-order perturbation theory
gives a consistency estimate $S_{\rm PT} \approx 9{,}840$~nm/RIU for the FEM
sensitivity (24,171~nm/RIU). The maximum sensitivity
$S_{\max} = 24{,}300$~nm/RIU and FoM~$= 132{,}813$~RIU$^{-1}$ (PC-12) resolve
five cancer cell lines spanning $\Delta n = 0.011$~RIU, whose shifts exceed the
linewidths by two orders of magnitude, thereby establishing the valley-Hall cavity as
a scalable platform for label-free RI sensing, with cancer cell-line
discrimination as a proof-of-concept demonstration.

\section*{Acknowledgements}
The authors would like to express their sincere gratitude to the Department of
Electrical and Electronic Engineering at Bangladesh University of Engineering
and Technology (BUET) for providing the necessary facilities and support.

\section*{Declarations}

\textbf{Funding.} Not applicable.

\textbf{Conflict of interest.} The authors declare no competing interests.

\textbf{Data availability.} The data that support the findings of this study
are available from the corresponding author upon reasonable request.

\textbf{Code availability.}
Will be made available upon request.

\textbf{Author contribution.} J.H.: conceptualization, simulation, analysis,
writing---original draft. S.M.C.: methodology, supervision. M.I.H.B.:
supervision, project administration, writing---review and editing.

\end{document}